\documentclass[aps,prb,twocolumn,showpacs,superscriptaddress, longbibliography, 10pt]{revtex4-1}

\usepackage{graphicx}
\usepackage{bm}
\usepackage{cleveref}
\usepackage{siunitx}
\usepackage{color}
\usepackage[normalem]{ulem}
\usepackage[version=4]{mhchem}
\bibpunct{[}{]}{,}{n}{}{}
\graphicspath{{figures/}}

\begin{document}

\title{Strain engineering a multiferroic monodomain in thin-film BiFeO$_3$}

 \author{N. Waterfield Price}
 \affiliation{Clarendon Laboratory, Department of Physics, University of Oxford, Parks Road, Oxford OX1 3PU, United Kingdom}

\author{A. M. Vibhakar}
\affiliation{Clarendon Laboratory, Department of Physics, University of Oxford, Parks Road, Oxford OX1 3PU, United Kingdom}

\author{R. D. Johnson}
\affiliation{Clarendon Laboratory, Department of Physics, University of Oxford, Parks Road, Oxford OX1 3PU, United Kingdom}
\affiliation{ISIS Facility, Rutherford Appleton Laboratory, Chilton, Didcot, OX11 0QX, United Kingdom}

\author{J. Schad}
\affiliation{Department of Materials Science and Engineering, University of Wisconsin-Madison, Madison, Wisconsin 53706, USA}

\author{\\W. Saenrang}
\altaffiliation[Present address: ]{School of Physics, Institute of Science, Suranaree University of Technology, Nakhon Ratchasima, 30000, Thailand}
\affiliation{Department of Materials Science and Engineering, University of Wisconsin-Madison, Madison, Wisconsin 53706, USA}

\author{A. Bombardi}
\affiliation{Diamond Light Source Ltd., Harwell Science and Innovation Campus, Didcot OX11 0DE, United Kingdom}

\author{F. P. Chmiel}
\affiliation{Clarendon Laboratory, Department of Physics, University of Oxford, Parks Road, Oxford OX1 3PU, United Kingdom}

\author{C.B. Eom}
\affiliation{Department of Materials Science and Engineering, University of Wisconsin-Madison, Madison, Wisconsin 53706, USA}

\author{P. G. Radaelli}
\email{paolo.radaelli@physics.ox.ac.uk}
\affiliation{Clarendon Laboratory, Department of Physics, University of Oxford, Parks Road, Oxford OX1 3PU, United Kingdom}

\date{\today}

\begin{abstract}
The presence of domains in ferroic materials can negatively affect their macroscopic properties and hence their usefulness in device applications. From an experimental perspective, measuring materials comprising multiple domains can complicate the interpretation of material properties and their underlying mechanisms. In general, \ce{BiFeO3} films tend to grow with multiple magnetic domains and often contain multiple ferroelectric and ferroelastic domain variants. By growing (111)-oriented \ce{BiFeO3} films on an orthorhombic \ce{TbScO3} substrate, we are able to overcome this, and, by exploiting the magnetoelastic coupling between the magnetic and crystal structures, bias the growth of a given magnetic-, ferroelectric-, and structural-domain film. We further demonstrate the coupling of the magnetic structure to the ferroelectric polarisation by showing the magnetic polarity in this domain is inverted upon \SI{180}{\degree} ferroelectric switching.
\end{abstract}

\pacs{999}

\maketitle

\section{Introduction}

The coupling of spontaneous electric, magnetic, and elastic orders within single-phase materials, known as multiferroics, could provide a platform for a new generation of memory devices combining non-volatile storage with efficient switching mechanisms~\cite{cheong2007multiferroics, khomskii2009trend, fiebig2016evolution}.This unique technological promise can only be fulfilled by achieving an exquisite control over \emph{multiferroic domains}, which form as a direct consequence of spontaneous symmetry breaking. Furthermore, from an experimental perspective the intrinsic properties of a prototypical device can be obscured by the presence of multiple domains, and would ideally be demonstrated within a single-domain architecture. Although exploiting particular combinations of domains can be advantageous~\cite{martin2008nanoscale, seidel2009conduction, heron2011electric}, their population and the exact location of the domain boundaries must be controlled with great precision to realise the full potential of multiferroic materials.

\begin{figure}[!ht]
    \includegraphics[width=0.46\textwidth]{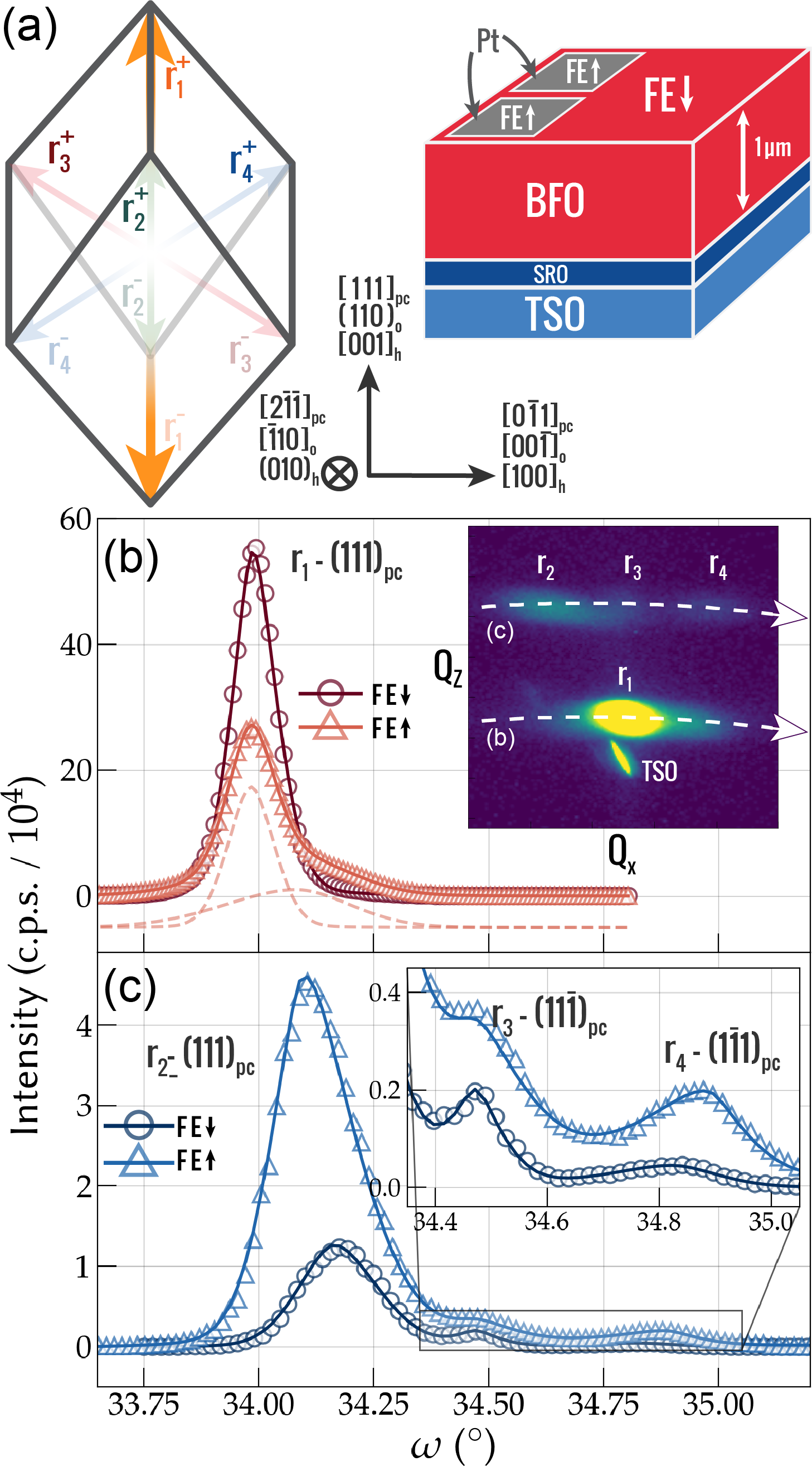}
    \caption{\textbf{Effect of switching on ferroelastic domains of BFO/TSO.} (a) Schematic depiction of the unique $[111]_\mathrm{pc}$ axes in each of the four ferroelastic domains, shown in the pseudocubic unit cell viewed along the $\mathbf{b}^*_\mathrm{h}$ direction. The $[111]_\mathrm{pc}$ axes are labelled r$_\text{1--4}$ with the $+/-$ superscript denoting the corresponding ferroelectric domain (see text). The crystallographic relationship between the hexagonal and pseudocubic BFO unit cells and the orthorhombic TSO unit cell is indicated by the black axes. Rocking curve scans of the (b) $(111)_\mathrm{pc}$ reflection of the majority r$_1$ ferroelastic domain and (c) $(\bar{1}11)_\mathrm{pc}$, $(11\bar{1})_\mathrm{pc}$, and $(1\bar{1}1)_\mathrm{pc}$ reflections corresponding to each of the minority r$_2$, r$_3$ and r$_4$ domains, respectively. Data for the as-grown FE$\downarrow$ and switched FE$\uparrow$ states are shown by the dark circles and light triangles, respectively. Fits to the data are shown by the solid lines and the dotted light-red lines in (b) show the two peak profiles contributing to the intensity in the FE$\uparrow$ state. The reciprocal space map shown in the inset is based on laboratory data (see Supplemental Material S-I~\cite{suppl} for more details), on which we indicated the position of the scans in (b) and (c).}\label{fig:characterisation}
\end{figure}

A classic illustration of this challenge is provided by \ce{BiFeO3} (BFO), one of the most studied multiferroics ~\cite{sando2014bifeo3}. At room temperature, BFO is at the same time ferroelectric, ferroelastic and antiferromagnetic (cycloidal)~\cite{kubel1990structure, sosnowska1982spiral}. Below the ferroelectric phase transition $T_\mathrm{C} \sim \SI{1100}{\kelvin}$, BFO has a polar R3c perovskite structure with a rhombohedral distortion along the (pseudocubic) $[111]_\mathrm{pc}$ direction~\cite{michel1969atomic}. This allows for four possible ferroelastic domains, (labelled r$_{\text{1--4}}$ according to the notation of Streiffer \emph{et al.}~\cite{streiffer1998domain}), each corresponding to the distortion being along each of the $\langle 111\rangle_\mathrm{pc}$ pseudocubic body diagonals, as shown in \cref{fig:characterisation}(a). Within each of these ferroelastic domains, there are two possible ferroelectric domains, r$_i^+$ and r$_i^-$, with the polarisation $\mathbf{P}$ along the $\pm [111]_\mathrm{pc}$ directions~\cite{zavaliche2006multiferroic}. Magnetic ordering occurs below $T_\mathrm{N} \sim \SI{640}{\kelvin}$, where the \ce{Fe} spins order with a long-period cycloidal magnetic structure. This breaks three-fold rotational symmetry and, for each ferroelectric domain, results in three symmetry equivalent $\mathbf{k}$ domains, with magnetic propagation vectors $\mathbf{k}_1 = (\delta, \delta, 0)_\mathrm{h}$, $\mathbf{k}_2 = (\delta, -2 \delta, 0)_\mathrm{h}$, and $\mathbf{k}_3 = (-2\delta, \delta, 0)_\mathrm{h}$, where the subscript $\mathrm{h}$ denotes the hexagonal setting and $\delta \sim 0.0045$ at room-temperature in the bulk~\cite{sosnowska1982spiral}. In fact, it has been recently reported that the structure of both bulk~\cite{sosnowska2012monoclinic} and thin film~\cite{waterfieldprice2016coherent, kan2010effect} BFO is monoclinic with broken three-fold symmetry. In this case, these monoclinic domains are coupled one-to-one to the magnetic $\mathbf{k}$-domains, resulting in three possible \emph{magnetostructural} domains. Each of these magnetostructural domains can support two possible directions of the magnetic polarity $\boldsymbol{\lambda}$, which is defined as $\boldsymbol{\lambda} = \mathbf{k} \times (\mathbf{S}_i \times \mathbf{S}_j)$, where $\mathbf{S}_i$ and $\mathbf{S}_j$ are spins on adjacent sites along the direction of $\mathbf{k}$, and can be thought of as characterising the ``sense of rotation'' of the cycloidal structure. It is understood that the magnetic polarity is directly coupled to the ferroelectric polarisation via the direct Dzyaloshinskii--Moriya interaction~\cite{johnson2013x}, giving combined ferroelectric/magnetopolar domains. In summary there are:
\vspace{-0.2cm}
\begin{equation*}
\overbrace{N_\mathrm{Ferroelastic}}^{4} \times \overbrace{N_\mathrm{Ferroelectric/Magnetopolar}}^{2} \times \overbrace{N_\mathrm{Magnetostructural}}^{3}
\end{equation*}
 $= 24$ possible domain variants.

In epitaxial films, the relative proportion of these domains can be manipulated. For instance, the ferroelastic domain populations of BFO may be tailored through certain choices of film orientation, substrate symmetry and miscut~\cite{giencke2014tailoring}. BFO films grown in the $(001)_\mathrm{pc}$ orientation on cubic \ce{SrTiO3} (STO) substrates, typically display all four ferroelastic domains. The number of ferroelastic variants can be reduced to two by growing BFO on a lower symmetry substrate, for example orthorhombic $(001)_\mathrm{pc}$-\ce{TbScO3} (TSO)~\cite{folkman2009stripe}. Alternatively, using a $(001)_\mathrm{pc}$-STO substrate miscut by \SI{4}{\degree} towards the $[100]_\mathrm{pc}$ direction favours the two ferroelastic domains with their rhombohedral distortion along the miscut direction~\cite{jang2009domain} and, similarly, a miscut towards $[110]_\mathrm{pc}$ instead results in a single ferroelastic domain~\cite{giencke2014tailoring}. A ferroelastic monodomain may also be achieved by growing a $(111)_\mathrm{pc}$-oriented film on an STO substrate, results in a single ferroelastic domain, with its rhombohedral distortion aligned with the $(111)_\mathrm{pc}$ surface-normal direction~\cite{kan2010effect}. The ferroelectric domain population can be controlled by applying an electric field or by including a metallic \ce{SrRuO3} buffer layer between the substrate and the BFO film, which screens the depoling field of the BFO and results in the electrical polarisation favouring directions with a downward component (into the film)~\cite{fong2006stabilization}. Although controlling the ferroelectric domain populations at growth is relatively straightforward, BFO-based ferroelectric devices are plagued by polarisation fatigue and electrical breakdown problems~\cite{baek2011nature}. This has been attributed to a nondeterministic, multistage switching process that causes the film to break up into multiple ferroelastic domains resulting in charged domain walls and domain pinning~\cite{zou2012mechanism}. Even though they can be grown as a ferroelectric monodomain, this issue is most pronounced in the $(111)_\mathrm{pc}$-oriented films, as this orientation preserves the intrinsic three-fold symmetry of the crystal structure, making all possible switching pathways equally likely.

By contrast, antiferromagnetic domains are more difficult to manipulate, because, unlike ferromagnets and ferroelastics, antiferromagnets do not couple to a uniform conjugated field.  Yet, achieving this control is crucial, because most device configurations include a ferromagnetic overlayer that is exchange-coupled to antiferromagnetic spins.  Although some control of propagation vectors using electric field~\cite{lee2008single} and high magnetic fields~\cite{bordacs2018magnetic} has been demonstrated, a robust method to bias a single propagation vector has yet to be found. Previously, single-magnetic-domain BFO had only been achieved in ultrathin, $(001)_\mathrm{pc}$-oriented films grown on a \ce{NdGaO3} substrate~\cite{kuo2016single} where films above a threshold thickness of $\sim \SI{5}{\nano\metre}$ were found to revert to a multidomain state. The recent discovery of a very strong magneto-elastic coupling~\cite{waterfieldprice2016coherent} in BFO affords a potentially new ``handle'' on magnetic domains. For example, we have previously shown that $(001)_\mathrm{pc}$-BFO films on miscut STO consist of a single antiferromagnetic domain~\cite{saenrang2017deterministic}, because the propagation vector oriented in the plane of the film is favoured over the other two. By contrast, for $(111)_\mathrm{pc}$-oriented BFO films grown on STO, the high symmetry of the substrate makes magnetic domain control extremely challenging. Typical $(111)_\mathrm{pc}$-oriented BFO films consist of a mosaic of magnetoelastic domains at the sub-micron scale, in which the three magnetic domains are equally represented~\cite{waterfieldprice2016coherent}.

In this Article, we demonstrate the ability to bias the growth of a given structural, ferroelastic, ferroelectric, and magnetic domain, in a \SI{1}{\micro\metre}-thick $(111)_\mathrm{pc}$-oriented BFO film by breaking three-fold symmetry with an orthorhombic TSO substrate. Remarkably, the substrate is able to bias the growth of a majority magneto-structural domain, in spite of the fact that most of the film is relaxed away from the substrate lattice parameters. After multiple switches of the ferroelectric polarisation, the majority of the sample (about \SI{80}{\percent}) remains in the same ferroelastic and magnetoelastic state, while about \SI{19}{\percent} of the sample converts to a specific alternate ferroelastic domain as a result of a more deterministic switching mechanism. Nonresonant x-ray magnetic scattering (NXMS) with polarisation analysis was employed to determine the magnetic polarity of the film and to observe its inversion after ferroelectric switching.

\section{Experimental details}

\SI{1}{\micro\metre}-thick epitaxial films of $(111)_\mathrm{pc}$-oriented BFO films were grown using double-gun off-axis sputtering~\cite{brewer2017uniform} onto a TSO substrate oriented with the $(110)_\mathrm{o}$ direction specular (the subscript o indicates the $Pnma$ orthorhombic setting). This direction approximates the $(111)_\mathrm{pc}$ direction of a simple perovskite, but has an in-plane anisotropy due to the orthorhombicity.  The BFO layer grows such that the $\mathbf{a}_\mathrm{h}$ axis is antiparallel to the $\mathbf{c}_\mathrm{o}$ axis of the TSO substrate. Before depositing the BFO layer, a \SI{30}{\nano\metre}-thick \ce{SrRuO3} (SRO) layer was deposited on the TSO substrate by \SI{90}{\degree} off-axis sputtering. The role of this layer is twofold: firstly, it elastically biases the film to grow as a ferroelectric monodomain~\cite{fong2006stabilization}, and secondly, it acts as a bottom electrode. After the BFO was grown, $\SI{500}{\micro\metre} \times \SI{200}{\micro\metre} \times \SI{10}{\nano\metre}$ \ce{Pt} top electrodes were patterned on the film surface using photolithography. This setup allows switching of the ferroelectric polarisation under the electrodes along the $(111)_\mathrm{pc}$ direction (out-of-plane). \SI{1}{\micro\metre}-thick, $(111)_\mathrm{pc}$-oriented BFO films where similarly grown on $(111)_\mathrm{pc}$-oriented STO substrates, details of which can be found in \cite{waterfieldprice2016coherent}. The films grown both on TSO and STO substrates show excellent ferroelectric characteristics (see Supplemental Material S-III~\cite{suppl}) with remnant polarisations of $\SI{108}{\micro\coulomb\per\centi\metre}$ and $\SI{110}{\micro\coulomb\per\centi\metre}$, respectively, comparable to the highest values reported for BFO films~\cite{das2006synthesis}.

The synchrotron x-ray measurements, including both the structural characterisation and nonresonant x-ray magnetic scattering (NXMS) measurements, were performed on beamline I16, Diamond Light Source (UK) using a six-circle kappa diffractometer in reflection geometry.  Additional measurements were performed on a Rigaku SmartLab four-circle laboratory-based x-ray diffractometer fitted with a copper K$_\alpha$ x-ray source (see Supplemental Material S-I).  For both structural characterisation and NXMS measurements, the incident synchrotron x-ray beam energy was tuned to \SI{4.9}{\kilo\electronvolt}, which is off-resonance of all chemical elements present in the sample. This energy is high enough that air scattering is minimal, while the NXMS signal can still be relatively easily separated from multiple scattering processes.  Conversion of the incident x-ray polarisation from linear to circular was achieved using a \SI{100}{\micro\meter}-thick diamond quarter-wave plate. For the NXMS measurements, the scattered x-ray polarisation was determined using a pyrolytic-graphite polarisation-analyser crystal, scattering at the (004) reflection. The x-ray beam size at the sample was adjusted by means of slits from $\sim \SI{300}{\micro\metre} \times \SI{50}{\micro\metre}$ to $\sim \SI{50}{\micro\metre} \times \SI{50}{\micro\metre}$ to allow a variable area of the sample, including a small spot under the electrodes, to be illuminated. For all the NXMS measurements, we used an avalanche photodiode detector, positioned after the polarisation-analyser crystal, while the structural characterisation measurements were performed using a Pilatus photon-counting area detector. Single-crystal neutron diffraction measurements were performed on the WISH instrument at ISIS, the UK pulsed Neutron and Muon Spallation Source. The neutron beam size at the sample position was $\sim \SI{20}{\milli\metre} \times \SI{40}{\milli\metre}$, thus illuminating the entire film. All the NXMS and neutron diffraction measurements were taken at room temperature.

\section{Results and Discussion}

\subsection{Structural and magnetic characterisation of the unswitched sample}

\begin{figure}[ht]
    \includegraphics[width=0.47\textwidth]{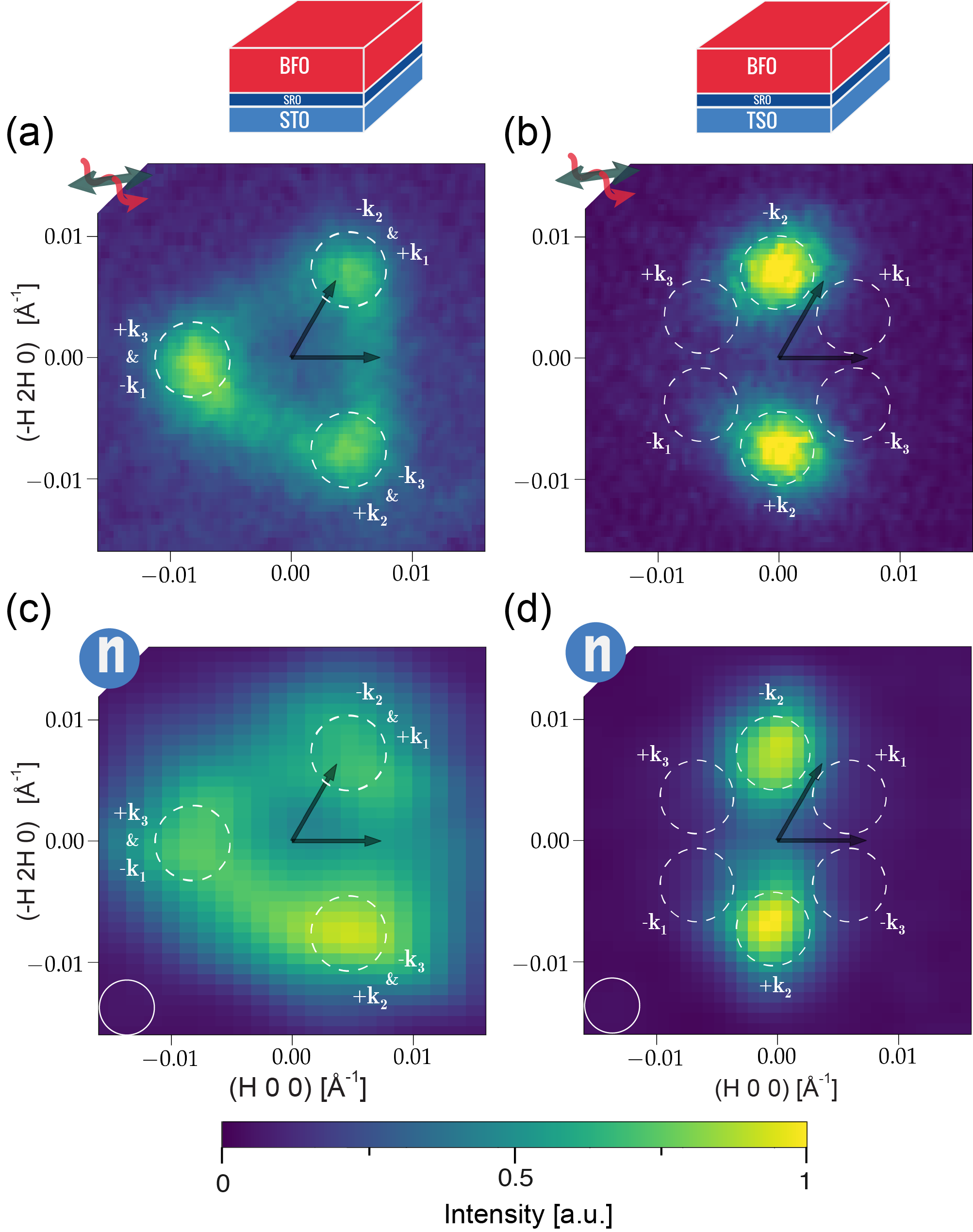}
    \caption{\textbf{Magnetic x-ray and neutron diffraction reciprocal space maps of BFO/STO and BFO/TSO.} Reciprocal space maps about the $(009)_\mathrm{h}$ position measured by NXMS with linearly-polarised incident x-rays for (a) BFO/STO and (b) BFO/TSO. Reciprocal space maps about the $(003)_\mathrm{h}$ position measured by neutron diffraction for (c) BFO/STO and (d) BFO/TSO. The BFO/STO neutron data in (c) are reproduced from \cite{waterfieldprice2016coherent} and x-ray data in (a) were measured on the same sample using the method described in \cite{waterfieldprice2016coherent}. The expected positions of the magnetic satellite peaks for the two different substrates are indicated by the dashed circles. The white circle in the lower-left corner of (c) and (d) indicates the instrumental resolution of the neutron measurements (the x-ray resolution is smaller than the pixel size). Due to the lower sensitivity of the single-crystal neutron measurements to the lattice parameters, the neutron data have been scaled to the lattice parameters determined from the higher-resolution x-ray measurements. The reciprocal lattice directions (in the hexagonal setting) are indicated by the translucent black arrows. All measurements were taken at room temperature.}\label{fig:rsms}
\end{figure}

The ferroelastic domain population of the unswitched sample (FE$\downarrow$) was determined from reciprocal space scans measured on beamline I16, in which 2-dimensional images from the Pilatus detector were collected as a function of the rocking angle $\omega$.  Intensities integrated along particular lines in reciprocal space (corresponding to rocking curves) are shown in \cref{fig:characterisation}(b, c), while corresponding 2D laboratory data are reported in Supplemental Material S-I~\cite{suppl}. The four possible ferroelastic domains are labelled r$_\text{1--4}$, as shown in \cref{fig:characterisation}(a). The r$_1$ domain has its rhombohedral axis aligned perpendicular to the film, so that the $(111)_\mathrm{pc}$ reflection is surface-normal. The $(\bar{1}11)_\mathrm{pc}$,  $(\bar1\bar{1}1)_\mathrm{pc}$ and $(11\bar{1})_\mathrm{pc}$ reflections of the r$_\text{2-4}$ domains are nearby in reciprocal space but have a slightly shorter $d$-spacing and form a triangle around the $(hhh)_\mathrm{pc}$ axis of the r$_1$ domain, indicative of coherent twinning. Note that, by the geometry of the measurement, this triangle is projected into a single arc in the reciprocal space map shown in the inset of \cref{fig:characterisation}(b). Almost \SI{96}{\percent} of the sample was found to be in the r$_1$ domain, which has a narrow rocking curve (\cref{fig:characterisation}(b)) with $\Delta\omega_\mathrm{FWHM} \sim$ 0.1$^\circ$.  The r$_2$ population was $\sim \SI{3.7}{\percent}$, which is significantly greater than for films grown on STO.  The rocking curves of the minority domains r$_\text{2--4}$ (\cref{fig:characterisation}(c)) were somewhat broader ($\Delta\omega_\mathrm{FWHM} \sim$ 0.2$^\circ$) than for r$_1$.

\begin{table}[b]
\centering
\caption{Ferroelastic domain fractions for the as-grown FE$\downarrow$ and switched, FE$\uparrow$ state, obtained from least-squares regression to data shown in \cref{fig:characterisation}.}
\label{tab:domains}
    \begin{ruledtabular}
    \begin{tabular}{l l l l l l}
    & \multicolumn{5}{c}{Domain populations (\%)}\\[0.5ex]
    State & r$_1$ & r$_1^\prime$ & r$_2$  & r$_3$ & r$_4$ \\[0.5ex]
    \hline\\[-2ex]
    FE$\downarrow$ & 95.77(2) & - & 3.77(7) & 0.31(2) & 0.14(1)  \\[0.5ex]
    FE$\uparrow$ & 46.73(41) & 33.61(100) & 18.12(16) & 0.77(5) & 0.76(4) \\[0.5ex]
    \end{tabular}
    \end{ruledtabular}
\end{table}

Next, we probed the magnetic structure of the majority (r$_1$) ferroelastic domain by NXMS and neutron diffraction. We measured the magnetic satellite reflections corresponding to the long-range incommensurate magnetic ordering, which occur near the $\mathbf{N} = (0, 0, 3(2n + 1))_\mathrm{h}$ reciprocal lattice positions at which (Thomson) charge scattering is forbidden due to the c-glide.

The diffraction pattern expected from each of the magnetic $\mathbf{k}$-domains is a pair of peaks at positions $\mathbf{N}\pm\mathbf{k}_i$, where $\mathbf{k}_i$ is the propagation vector of the corresponding magnetic domain. Hence, if populations of all three magnetic domains are illuminated, one should observe a star of six satellite peaks. For $(111)_\mathrm{pc}$-oriented BFO films grown on STO, we have previously shown that these six peak merge in pairs to form a triangle of peaks, due to the presence of magnetostructural domains on a sub-micron scale~\cite{waterfieldprice2016coherent} (NXMS and neutron diffraction data from STO-grown samples are reproduced in \cref{fig:rsms}(a) and (c)).  \Cref{fig:rsms}(b) shows a reciprocal space map of the TSO-grown sample measured by NXMS about the $\mathbf{N} = (009)_\mathrm{h}$ reciprocal lattice position measured in a region of the film in its as-grown `virgin' state. These data were collected using linearly-polarised incident light with $\mathbf{E}$ normal to the scattering plane whilst only detecting scattered light with $\mathbf{E}$ in the scattering plane (polarisation analyser angle of $\eta = \SI{90}{\degree}$). In sharp contrast with the BFO/STO data, we observe only a \emph{single pair of satellite peaks} at positions $\mathbf{N} \pm \mathbf{k}_2$. Therefore, in the region of the film illuminated by the x-ray spot, r$_1$ ferroelastic domains support a single magnetic domain. We repeated this measurement at several sample locations over two identically grown BFO/TSO films and obtained identical results at all positions. To further test this magnetoelastic coupling, we measured an equivalent, unswitched film using neutron diffraction, and the corresponding reciprocal space map about the $(003)_\mathrm{h}$ position is shown in \cref{fig:rsms}(d). Given that the neutron beam illuminates the entire sample, this verifies that all r$_1$ domains comprise a magnetic monodomain. Fitting a theoretical intensity calculation to both the x-ray and neutron data simultaneously yields a propagation vector of $\delta = \SI{0.0042(1)}{\per\angstrom}$. The propagation vector of the cycloidal domain is $\mathbf{k}_2$, which is aligned along the $[001]_\mathrm{o}$ direction of the TSO substrate. We also note the absence of any diffuse tails of intensity extending between the two satellite peaks and the $(009)_\mathrm{h}$ position, as is seen in $(111)_\mathrm{pc}$-oriented BFO films grown on STO substrates~\cite{waterfieldprice2016coherent}. This indicates that the magnetic structure is largely coherent throughout the film and is consistent with the hypothesis that the small domain size in the multidomain BFO/STO films is an important factor in the disorder resulting in the diffuse scattering observed in those films. Given that these films are grown as FE monodomains, one expects that the magnetic structure constitutes a monodomain not only in terms of having a single propagation vector, but also a single magnetic polarity. The magnetic polarity is further explored in \cref{sec:NXMS}. Although time-inversion/phase-slip domains may still be present, these would not affect the ability of the BFO film to exchange-couple to a ferromagnetic overlayer.

\subsection{Structural and magnetic characterisation of the switched sample}

Additional structural and magnetic diffraction data were collected under an electrode, which was electrically switched a single half-cycle of the ferroelectric hysteresis loop (FE$\uparrow$).  Rocking curves of the four ferroelastic domains are reported in \cref{fig:characterisation}(b, c) with the same conventions as for the unswitched sample.  There are two main effects to be remarked:  firstly, part of the intensity is transferred from the r$_1$ to the r$_2$ domain, and, secondly, the $(111)_\mathrm{pc}$ reflection of the r$_1$ domain broadens. After switching, the r$_1$ intensity is well modelled with a sharp peak (which we continue to label r$_1$) centred at the same position, and with a similar width to the unswitched peak but approximately half the intensity, plus a broader peak, which we will refer to as r$_1^\prime$, shifted $\sim$ 0.1$^\circ$ to higher $\omega$. The presence of this broad peak indicates that a significant fraction of the sample has become misaligned with respect to the surface normal during the switching process. Both the bulk ferroelectric measurements (see Supplemental Material S-I) and the NMXS polarimetry data (see \cref{sec:NXMS}), indicate that the electric polarisation has been switched in most of the sample under the electrode.

The relative ferroelastic domain populations for the unswitched and switched states are listed in \cref{tab:domains}, and we see that over half the population of the r$_1$ domain has been lost, most of which has moved into r$_1^\prime$ and r$_2$, and that domain populations of r$_3$ and r$_4$ have increased only slightly. The skewed populations of the r$_\text{2--4}$ domains after switching has important implications for the FE switching mechanism. It has been demonstrated that \SI{180}{\degree} switching in $(111)_\mathrm{pc}$-BFO films occurs via a stochastic, multi-step switching path consisting of three \SI{71}{\degree} switching events via the intermediate FE/ferroelastic states~\cite{baek2011nature}. In said study, Baek \emph{et al.} report that, after many switching cycles of $(111)_\mathrm{pc}$-BFO/STO, they observe equal populations of the three intermediate ferroelastic domains. This is consistent with the fact that three non-surface-normal BFO ferroelastic domains are symmetry equivalent by the three-fold axis of the $(111)_\mathrm{pc}$-STO substrate. In the present case, the TSO substrate has imposed a \emph{preferential switching pathway} through the r$_2$ domain. This also follows naturally from symmetry considerations, as the orientation of the r$_\text{2-4}$ domains with respect to the orthorhombic $\mathbf{c}_\mathrm{o}$ axis breaks the symmetry between r$_2$ (tilted perpendicularly to $\mathbf{c}_\mathrm{o}$) and r$_3$/r$_4$ (tilted along directions containing equal and opposite components parallel to $\mathbf{c}_\mathrm{o}$).

The magnetic reflections of the switched state were found to be significantly broader (see Supplemental Material S-III), consistent with the reduced crystalline quality of the film upon switching i.e. size effects from smaller ferroelastic domains and the introduction of structural disorder.

\begin{figure*}[t]
    \includegraphics[width=\textwidth]{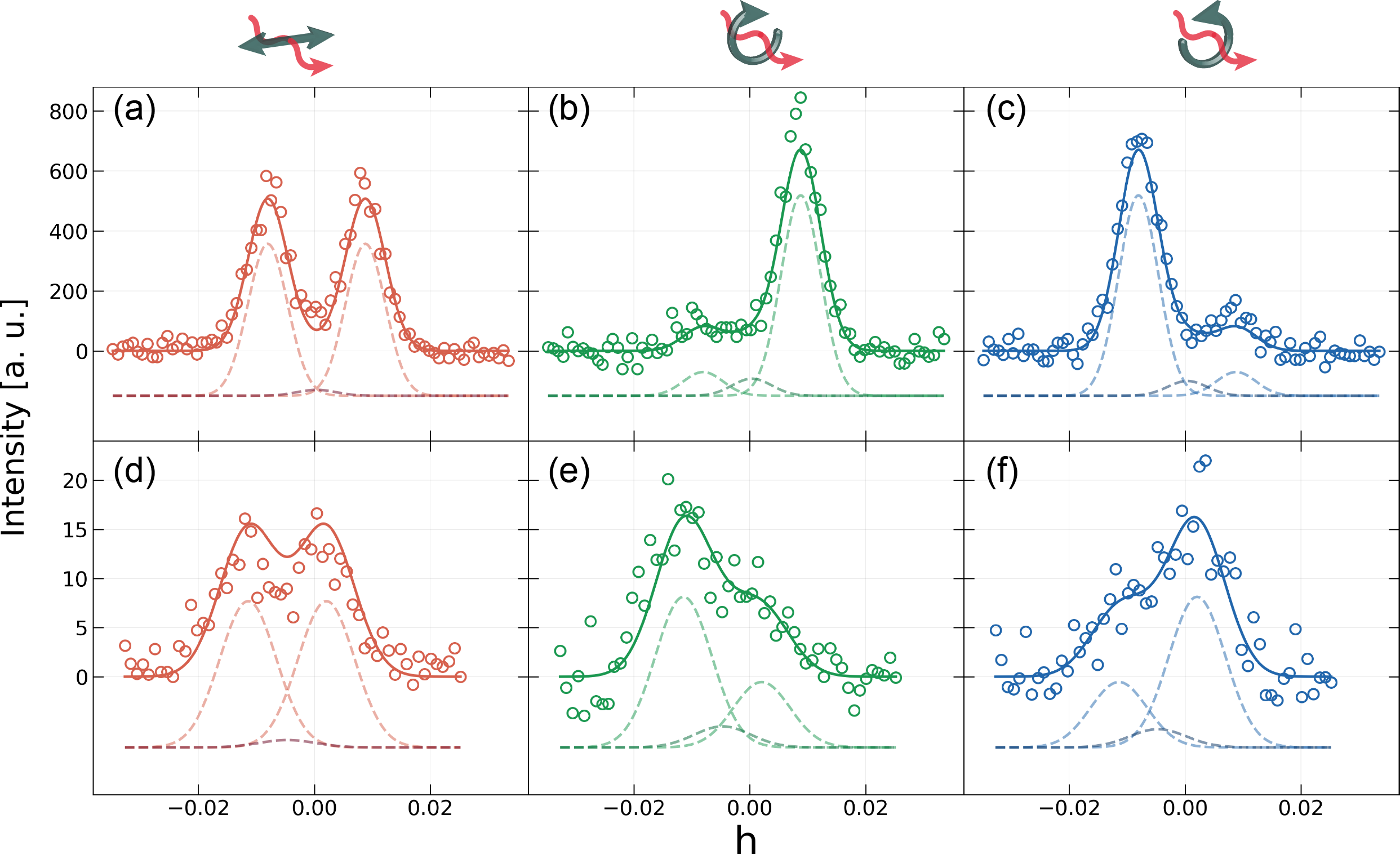}
    \caption{\textbf{Polarisation dependence of the NXMS intensity of the $\pm \mathbf{k}_2$ satellite peaks.} Reciprocal space scans through the $(009)_\mathrm{h} \pm \mathbf{k}_2$  peaks of a BFO/TSO film measured in the unswitched (a--c) and switched (d--f) states with incident linear horizontal (left), circular left (middle) and circular right (right) polarised x-rays, as indicated. The circles show the experimentally measured intensities and the solid lines show the results of a least-squares refinement of the scattered intensity for a single cycloidal domain. All measurements were taken at room temperature.}\label{fig:eta_scans}.
\end{figure*}

\subsection{NMXS polarimetry}\label{sec:NXMS}

To further probe the magnetic structure of the BFO/TSO monodomain film, we measured the polarisation dependence of the magnetic scattering.  Here, our approach is very similar to that employed for the STO sample~\cite{waterfieldprice2016coherent}.  Using linear horizontal and left/right circular polarised incident x-rays, we measured reciprocal space scans through both $\mathbf{N} \pm \mathbf{k}_2$ magnetic satellite peaks as a function of polarisation analyser angle. These measurements are highly sensitive to both the plane of rotation and magnetic polarity of the cycloidal magnetic structure~\cite{waterfieldprice2016coherent}. A model parameterised in terms of the cycloid plane, propagation vector, magnetic polarity, a small multiple scattering peak at the $(009)_\mathrm{h}$ position, peak parameters, a (polarisation dependent) constant background and a depolarisation factor, was fit to the data (see Supplemental Material S-II~\cite{suppl} for more details).

\Cref{fig:eta_scans}a-c show selected reciprocal space scans measured with a polarisation analyser angle of $\eta = \SI{90}{\degree}$, with the result of the above least-squares refinement shown as solid lines. The theoretical intensity ratio of the two diffraction peaks, for incident circularly polarised x-rays and at this polarisation analyser angle, is given by (see Supplemental Material S-II~\cite{suppl})

\begin{equation}
\frac{I_\mathrm{+\mathbf{k}_2}}{I_\mathrm{-\mathbf{k}_2}} \approx \frac{1 + \beta\gamma}{1 - \beta\gamma}
\label{eqn:intensity_ratio}
\end{equation}

where the $\gamma = +1$ or $-1$ for left-circular and right-circular polarised light, respectively. The sign of the magnetic polarity is parameterised by $\beta = -[(\boldsymbol{\lambda} \cdot \hat{\mathbf{n}})/|\boldsymbol{\lambda}|]$, where $\hat{\mathbf{n}}$ is the film surface normal. In agreement with previous studies both on bulk single crystals~\cite{johnson2013x} and films (grown on STO substrates)~\cite{waterfieldprice2017electrical}, we find experimentally that $\beta = -1$ for the FE\,$\downarrow$ state, such that the magnetic polarity is aligned antiparallel to the FE polarization.

The model fits the data extremely well, with a reduced chi-squared statistic of $\chi_\nu = 1.24$ and we thus find that the data is most consistent with a cycloidal magnetic structure with spins rotating in a plane containing the $[001]_\mathrm{h}$ (out-of-plane) direction and the $\mathbf{k}_2$ (in-plane) propagation vector direction, i.e., the same magnetic structure as for the bulk \cite{sosnowska1982spiral, johnson2013x}. These scans further corroborate that the r$_1$ domain is a monodomain in the magnetic polarity, as the opposite magnetic polarity would have the opposite circular polarisation dependence.

To test the response of the magnetic structure to \SI{180}{\degree} electrical switching within the majority r$_1$/r$_1^\prime$ domain~\footnote{Note that we are primarily sensitive to the r$_1$ domain, but, as can be seen from \cref{fig:characterisation}(b), there will be some finite contribution from r$_1^\prime$ scattering to the same point in reciprocal space.}, we measured the NXMS signal from regions of the film that have been switched into the FE$\uparrow$ state. \Cref{fig:eta_scans}d--f show, for a region of the film switched into the FE$\uparrow$ state, reciprocal space scans through the $\mathbf{N} \pm \mathbf{k}_2$ magnetic satellite peaks measured with incident linear horizontal, circular left, circular right polarised x-rays, respectively.  It is clear that the polarisation dependence of the scattered intensity for left-/right-circular polarised x-rays is inverted between the two FE polarisation states and, as expected, the linear polarisation dependence remains the same. A simultaneous fit of the polarisation analyser scans including both the switched and unswitched data demonstrated a \SI{79(4)}{\percent} switch of the magnetic polarity from the as-grown FE$\downarrow$ state to the switched FE$\uparrow$ state. When performing an equivalent measurement on a multi-$\mathbf{k}$-domain $(111)_\mathrm{pc}$-BFO/STO film, we observed a \SI{93(1)}{\percent} switch of the magnetic polarity upon inverting the ferroelectric polarisation~\cite{waterfieldprice2017electrical}. In both cases, it is possible that the unswitched fraction is due to pinned FE$\downarrow$ domains at the SRO/BFO interface, or that some of the film relaxed back in to the FE$\downarrow$ state, and the magnitude of these effects was enhanced by the TSO substrate.

\section{Conclusion}

In summary, by using anisotropic epitaxial strain from an orthorhombic TSO substrate in the growth of $(111)_\mathrm{pc}$-oriented BFO films, we found that we were able to bias the growth of a majority ferroelastic domain. We then demonstrated that this ferroelastic domain supported a single magnetic cycloidal $\mathbf{k}$-domain. Using x-ray polarimetry, we were able to confirm that this magnetic domain was also of a single cycloidal magnetic polarity. As for previous samples grown on STO substrates~\cite{waterfieldprice2017electrical}, we were able to locally address a prototypical device structure and found that the magnetic polarity is inverted upon reversal of the FE polarisation. Finally, we found that, by using the orthorhombic TSO substrate, we were able to bias a preferential FE switching pathway, a critical ingredient if one is to achieve deterministic operation of a BFO-based device.

\section*{Acknowledgements}
The authors would like to thank Pascal Manuel for assistance with the neutron diffraction measurements. We acknowledge Diamond Light Source for time on Beam line I16 under Proposals MT15087-1 and MT17291-1. The work done at the University of Oxford (N.W.P., A.M.V., R.D.J., F.P.C., and P.G.R.) was funded by EPSRC grant No. EP/M020517/1, entitled “Oxford Quantum Materials Platform Grant.” The work at University of Wisconsin-Madison (W.S., J.S., and C.B.E) was supported by the Army Research Office through grant W911NF-13-1-0486. R.D.J. acknowledges support from a Royal Society University Research Fellowship.

\bibliography{references}

\end{document}